\documentclass[prl,amssymb,amsmath,twocolumn]{revtex4}

\begin{document}

\title{
Dynamical and scaling properties of $\nu=\frac{5}{2}$ interferometer
}
\author{Bas J. Overbosch}
\author{Xiao-Gang Wen}
\affiliation{Department of Physics, Massachusetts Institute of Technology, Cambridge, Massachusetts 02139, USA}
\date{\today}

\begin{abstract}
We calculate the non-linear I-V tunneling curves for a two-point-contact
tunneling junction between two edges of the $\nu=\frac{5}{2}$ non-abelian
fractional quantum Hall state.  The non-linear I-V tunneling curves are
calculated for both cases with and without an $e/4$ non-abelian
quasiparticle between
the two contacts.  We confirm that, within a dynamical edge theory, the
presence of the $e/4$ quasiparticle between the two contacts destroys the
interference between the two tunneling paths.
We also calculate how the interference reappears as the
$e/4$ quasiparticle is moved closer to an edge.
\end{abstract}

\maketitle

\section{Introduction}

The possibility of having anyons, particles in two dimensional systems which
obey fractional statistics in between bosons and fermions, is a remarkable
prediction by theory \cite{LM7701,W8257}. To observe anyons experimentally 
is hard however, as 
the signature of fractional statistics lies not in the
observation of anyonic thermodynamics, but rather in the observation of
Aharonov-Bohm type interference in quasiparticle motions.  

For abelian anyons, the Aharonov-Bohm interference of a quasiparticle is
described by a complex phase \cite{ASW8422}. Such a phase can be changed in
discrete steps when some extra quasiparticles are enclosed by the two
interference paths.  

However, for the more exotic non-abelian anyons, the Aharonov-Bohm
interference is described by matrices that act on an global Hilbert space of
topologically protected degenerate states \cite{Wnab,MR9162,BW9215}, which
represent a potentially decoherence free \cite{WNtop} qubit 
realization \cite{K032}.
Such an unusual fractional statistics is called non-abelian fractional
statistics.  For non-abelian anyons not only the phase but also the magnitude
of the interference can be affected. Complete destructive interference would
e.g. be a manifestation of non-abelian statistics \cite{SFN0502,SH0602,BKS0603,FFN0717}.

A candidate system that allows non-abelian anyons is the Moore-Read Pfaffian
description \cite{MR9162,GWW9105,Wnabhalf,FNTW9804} for the $\nu=\frac{5}{2}$ fractional
quantum Hall state in which interference would occur in the tunneling current
induced by quantum point contacts. For the Moore-Read state, a setup with
destructive interference has been predicted
theoretically \cite{PhysRevB.55.2331,SFN0502,SH0602,BKS0603,FFN0717}. In such a setup, an $e/4$
non-Abelian quasiparticle is trapped in a central island between two tunneling
contacts.  In the regime of weak tunneling the temperature and bias dependence
of the tunneling conductance can be calculated exactly through the edge state
theory \cite{Wnabhalf}.

The proposed interference in the $\nu=\frac{5}{2}$ state is destructive to
leading order in the point contact tunneling amplitudes. Higher order
contributions can potentially restore some interference. Here, we consider one
such higher order process. To be specific we allow for the exchange of the
charge-neutral fermionic quasiparticle in the $\nu=5/2$ 
Pfaffian state between the central island and one of the two edges.
Such a process gives rise to a non-vanishing interference pattern. This
contribution can be distinguished from leading order interference pattern
through the different scaling dependences on temperature and bias.

There are other processes which could potentially restore interference, for
example the tunneling of the island quasiparticle into one of the edges.
Such a process is rather complicated to describe theoretically however. 
Through a detailed numerical calculation, it was found that the
the neutral excitations on the edge have much lower energy scale
than the charged excitations.\cite{WYR0604} This suggests the possibility that
the neutral-fermion tunneling may dominate the $e/4$-quasiparticle tunneling.

The tunneling of the neutral fermionic quasiparticle also has the advantage
that it can be treated in the existing framework of leading order perturbative
calculation. Yet it touches on the fundamental properties of the non-abelian
anyon state as the neutral fermion tunneling flips the occupation of the
topologically protected zero mode.

\section{Interferometer for the Moore-Read state with destructive
interference}

\subsection{Tunneling conductance in FQH interferometer, average and amplitude}

\begin{figure*}
 \includegraphics[width=\textwidth]{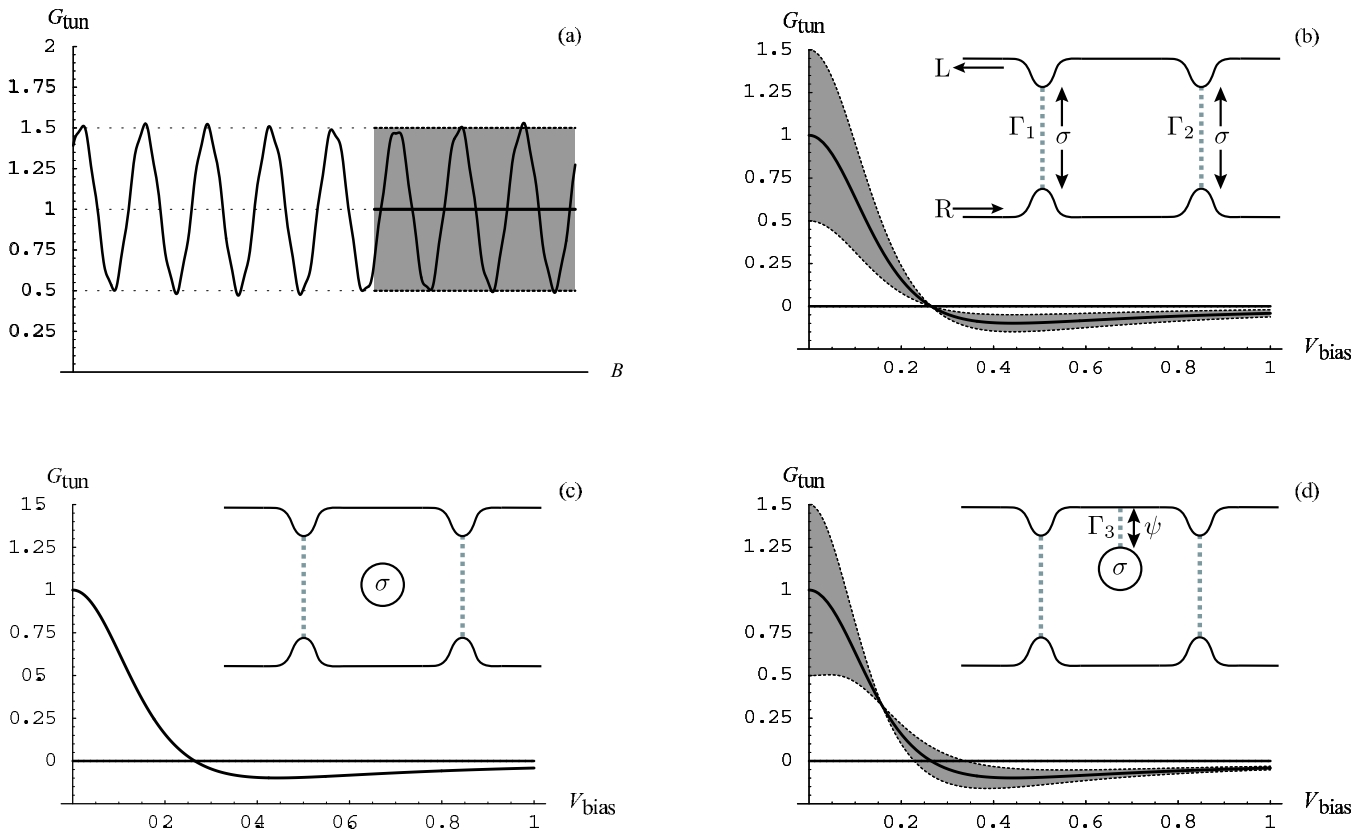}
\caption{
Sketch of main predictions for interference behavior in tunneling conductance
$G_\text{tun}$, for fixed temperature $T$ and in the weak-tunneling regime.
(a) At fixed bias voltage $V_\text{bias}$ interference oscillations may be
observed in $G_\text{tun}$ as a function of magnetic field $B$ (or any other
Aharonov-Bohm phase changing variable), and \emph{average} (thick line) and
\emph{oscillation amplitude} (grayed area) of the interference oscillations can be
determined.
(b) Behavior of average and amplitude of interference oscillations
of $G_\text{tun}$ as a function of $V_\text{bias}$ for interferometer with no
island. Both average and amplitude are given by $f_0(V_\text{bias})$.  
(c) For
setup with a $\sigma$-quasiparticle at the central island the interference in
$G_\text{tun}$ is completely deconstructive and fully vanishes, i.e., the
amplitude of oscillations is zero and only the average remains.  
(d) Interference is restored
when the central island with the $\sigma$-quasiparticle can have
$\psi$-$\psi$-tunneling to one of the edges. However, average and amplitude now
scale differently with bias, average like $f_0(V_\text{bias})$ and amplitude
like $f_1(V_\text{bias})$.  This difference in scaling can be used to
distinguish situations (b) and (d) when experimentally some interference is
observed; as a function of $T$ (b) and (d) also scale differently (not shown).
$V_\text{bias}$ is given in units of $2\pi T/q$ and units of $G_\text{tun}$
are arbitrary, with average and amplitude of $G_\text{tun}$ at zero bias set to
$1$ and $0.5$ respectively.
\label{fig:mainplot}
}
\end{figure*}

The interferometer concept that is used for generic quantum Hall fluids is
depicted in the inset of Fig.~\ref{fig:mainplot}b. Two edges, one left- and one right-moving, of a quantum hall
fluid are brought closer together by applying quantum point contacts; an
applied bias voltage between the two edges will then induce a tunneling
current of quasiparticles at these point contacts. In case of two or more
point contacts, the multiple ways in which a quasiparticle can tunnel can
cause interference of these paths; the presence of a non-trivial central
island between the point contacts can affect the interference due to the
fractional statistics of the quasiparticles.

The low-energy gapless modes are described by the appropriate chiral Luttinger
liquid edge state theory \cite{Wedgerev}. The best candidate to describe the
fractional quantum Hall state at filling fraction $\nu=\frac{5}{2}$ is
generally believed to be the Moore-Read Pfaffian state \cite{MR9162}; this
claim is supported by numerical simulations for closed
systems \cite{GWW9105,Wnabhalf,FNTW9804}, but has not yet been established
experimentally.

The edge states of the Moore-Read state are described by a free chiral charged
boson plus free chiral Majorana fermion theory \cite{Wnabhalf}. The
corresponding conformal field theory is a central charge $c=1$ plus a
$c=\frac{1}{2}$ theory. Each edge in the quantum Hall liquid carries a sector
label. Each sector forms an irreducible representation  of the algebra
of the electron operators. In CFT language electron operators are certain
primary fields of the $c=1+\frac{1}{2}$ theory.  The different sectors are
labeled by more general primary fields of the $c=1+\frac{1}{2}$ theory.  The
non-abelian part of the Moore-Read state comes from the $c=\frac{1}{2}$ Ising
contribution which carries a primary field $\sigma$ with fusion rules
$\sigma\times\sigma=\mathbbm{1}+\psi$, $\psi\times\psi=\mathbbm{1}$,
$\sigma\times\psi=\sigma$. The $c=1$ part is described by a bosonic field
$\phi$.

The allowed sectors for each edge are constrained only by the condition that
the fusion product of \emph{all} edges in the quantum Hall liquid combined is
in the identity representation. Otherwise we will consider the edges to be
fully independent of each other, such that we let operators on different edges
commute. We will consider the central island to be an edge as well, but with a
finite edge length described by a discrete set of (no longer gapless) modes; a
bulk quasiparticle can be considered as a zero-length island-edge that is
labeled by a sector only but has no additional modes.

The Hamiltonian we consider is a copy of the free edge theory for each edge
(see \cite{Wnabhalf}) plus tunneling operators which destroy a quasiparticle
on one edge and create it on the edge on the opposite side of a point contact.
As usual we restrict ourselves here to tunneling of the most relevant operator
with the smallest fractional charge, the charge $q=\frac{1}{4}$ quasiparticle
with operator $\sigma e^{\frac{i}{2}\phi}$ \cite{Wnabhalf}. With
$\omega_\text{J}=q e V$ the applied bias voltage between edges $\text{L}$ and
$\text{R}$, and $\Gamma_1$ and $\Gamma_2$ the tunneling amplitudes at the two
point contacts, the tunneling Hamiltonian becomes
\begin{align}
\label{eq:Htunneling}
H_\text{tun}(t)&=
\Gamma_1 e^{i\omega_\text{J} t}\sigma_\text{L}(t,x_1) e^{\frac{i}{2}
\phi_\text{L}(t,x_1)}\sigma_\text{R}(t,x_1) 
e^{-\frac{i}{2}\phi_\text{R}(t,x_1)}
\nonumber \\
+&\Gamma_2 e^{i\omega_\text{J} t}\sigma_\text{L}(t,x_2) 
e^{\frac{i}{2}\phi_\text{L}(t,x_2)}\sigma_\text{R}(t,x_2) 
e^{-\frac{i}{2}\phi_\text{R}(t,x_2)}
\nonumber\\
+& \text{H.c.}.
\end{align}

The tunneling current can be calculated in linear response to leading order in
the tunneling amplitudes $\Gamma_1$ and $\Gamma_2$ by expressing the tunneling
current in terms of a time-integral of time-ordered ground state correlation
functions of CFT primary fields, which are known exactly both at zero
temperature and finite temperature. We assume that the correlation function of
a product of operators on different edges factors out completely into a
product of correlation functions for each edge, and can be decoupled further
into free boson and free Majorana fermions correlators.  Following
Ref.~\cite{PhysRevB.55.2331} we find for the tunneling current 
\begin{gather}
I_\text{tun}(T,V)=|\Gamma_\text{eff}|^2 T^{-\frac{1}{2}}
F_0\!\left(\frac{qV}{2\pi T}\right)\\
|\Gamma_\text{eff}|^2=|\Gamma_1|^2+|\Gamma_2|^2+2\Re\!\left(\Gamma_1\Gamma_2^*e^{i\theta_\text{AB}}\right)\!H(V,T,\bar
x)\notag.
\end{gather} 
The generic structure of the tunneling current consists of three
parts: a dimensionful powerlaw temperature factor, here $T^{-1/2}$, which can
be read off directly from the CFT weights of the primary fields, a
dimensionless function $F_0(y)$ which gives the non-trivial dependence on bias
voltage (as determined by the primary field weights \footnote{
Explicit form for $F_0(y)$ can be read off from Eq.~(A25) in Ref.~\cite{PhysRevB.55.2331} with $g=\frac{1}{4}$.
}), and finally an effective
tunneling amplitude $|\Gamma_\text{eff}|^2$. The effective tunneling amplitude
is of the form typical for interference: a contribution from each tunneling
path separately, $|\Gamma_1|^2$ and $|\Gamma_2|^2$, plus a pure interference
term which includes the Aharonov-Bohm phase $\theta_\text{AB}$.

The zero temperature limit, $T\ll V$, is properly taken care of
as $F_0(V/T)\sim(V/T)^{-1/2}$ in this limit. The interference-term also includes a dimensionless function $H(V,t,\bar x)$
which incorporates the effect of the distance $\bar x$ between the two point
contacts. Because the typical experimental $\bar x$ is much smaller than $V$
and $T$ (when expressed in unit of length) and in this limit, $H(V,T,\bar
x=0)=1$, we assume we can ignore $H(T,V,\bar x)$ altogether for our purposes.

  Experimentally, it is not the
tunneling current itself which is measured in FQH interferometers, but rather
the differential tunneling conductance $G_\text{tun}=\partial I_\text{tun}/\partial V$
\footnote{Typically, it is the the
differential longitudinal resistance as function of bias current which is
measured experimentally using lock-in techniques, but at a quantum Hall
plateau this is equivalent to differential tunneling conductance as function
of bias voltage.}, and we
can write 
\begin{equation} 
G_\text{tun}(T,V)=|\Gamma_\text{eff}|^2
T^{-\frac{3}{2}} f_0\!\left(\frac{qV}{2\pi T}\right),
\end{equation}
where $f_0(y)=F_0'(y)$ and we absorbed the constant factors $q/2\pi$ into the tunneling amplitudes $\Gamma_i$.

The differential tunneling conductance has a form
\begin{gather}
 G_\text{tun}(T,V)=G^\text{ave}_\text{tun}(T,V)+G^\text{osc}_\text{tun}(T,V)\cos(\theta_\text{AB}+\theta_0),\notag\\
G^\text{ave}_\text{tun}(T,V)=\left(|\Gamma_1|^2+|\Gamma_2|^2\right)\,T^{-\frac{3}{2}}\,f_0\!\left(\frac{qV}{2\pi T}\right),
\label{eq:averageconductance}
\\
G^\text{osc}_\text{tun}(T,V)=2|\Gamma_1\Gamma_2|\,T^{-\frac{3}{2}}\,f_0\!\left(\frac{qV}{2\pi T}\right)\notag
.
\end{gather}
$G^\text{osc}_\text{tun}(T,V)$ describes the amplitude of the interference
oscillation, and $G^\text{ave}_\text{tun}(T,V)$ the average differential
tunneling conductance.  We note that both $G^\text{osc}_\text{tun}(T,V)$ and
$G^\text{ave}_\text{tun}(T,V)$ scale as $T^{-3/2}$.
In fact, they both depend on $V/T$ in the same way.

\subsection{Vanishing interference with $\sigma$ quasiparticle on central island}

The non-abelian statistics of the quasiparticles can be probed if the
interferometer contains a non-trivial edge, e.g., a small central island (or
bulk quasiparticle) in the $\sigma$ sector, as depicted in the inset of
Fig.~\ref{fig:mainplot}c \footnote{Quasiparticles in the $\sigma$ sector can
only exist in pairs, so we assume that there is another $\sigma$-quasiparticle
on another island in the FQH fluid outside the interferometer.}. Now a
non-trivial braiding, associated with a quasiparticle going around the path
that is enclosed by the interferometer, enters the interference.

In order to account for the braiding in the interference we have to pay close
attention to the fusion channels of our operators, guided by the following two
principles: 1) a tunneling event creates quasiparticles on two edges which are
in the \emph{identity} channels as tunneling cannot change the sector of the
total system, and 2) only operators which are in the identity channel on
\emph{each} edge have a non-vanishing expectation value. For abelian states
one already used these principles as they encompass charge conservation. What
non-abelian statistics adds to this is that for a given operator there can be
more than one fusion channel and braiding can affect the channel.

Notationwise it is difficult to capture fusion and braiding concisely due to
the inherent two-dimensional aspect of  braiding; our notation is clearly not
ideal but should suffice for our purposes.

We envision a product of operators which includes a tunneling operator of a
$\sigma$-quasiparticle between left and right edges at location $i$, which we
describe by $\big[\sigma_\text{L}\sigma_\text{R}\big]_i^\mathbbm{1}$, where
the superscript $\mathbbm{1}$ indicates that this tunneling operator is in the
identity channel. Furthermore there is a similar tunneling operator at
location $j$. The locations $i$ and $j$ and the Left and Right edges define an
enclosed Island which can be endowed with a sector of the theory.
If we let the sector of this island be the identity (as indicated by $(\mathbbm{1})_\text{I}$, i.e., there is no non-trivial
non-Abelian quasiparticle inside the enclosed island) 
we find the fusion of the two $\sigma_\text{L}$ and two $\sigma_\text{R}$
tunneling operators gives rise to
\begin{equation}
 \big[\sigma_\text{L}\sigma_\text{R}\big]_i^\mathbbm{1}\big[\sigma_\text{L}\sigma_\text{R}\big]^\mathbbm{1}_j\xrightarrow[\displaystyle (\mathbbm{1})_\text{I} ]{\text{fus.ch.}}\mathbbm{1}_\text{L}\mathbbm{1}_\text{R}\mathbbm{1}_\text{I}+\psi_\text{L}\psi_\text{R}\mathbbm{1}_\text{I},
\end{equation} 
indicating that there exist two possible fusion channels
including one that is the identity representation at all edges (left, right
and island edges).

If the enclosed region contains a $\sigma$ sector 
(i.e. a charge $e/4$ non-Abelian quasiparticle),
then 
the same tunneling operator $\big[\sigma_\text{L}\sigma_\text{R}\big]_i^\mathbbm{1}\big[\sigma_\text{L}\sigma_\text{R}\big]^\mathbbm{1}_j$
will fuse differently
\begin{equation}
\big[\sigma_\text{L}\sigma_\text{R}\big]_i^\mathbbm{1}\big[\sigma_\text{L}\sigma_\text{R}\big]^\mathbbm{1}_j\xrightarrow[\displaystyle(\sigma)_\text{I}]{\text{fus.ch.}}\psi_\text{L}\mathbbm{1}_\text{R}\psi_\text{I}+\mathbbm{1}_\text{L}\psi_\text{R}\psi_\text{I}.
\label{eq:sigmafusionchannel}
\end{equation} 
Note that if we move a charge $e/4$ non-Abelian quasiparticle around another
$e/4$ non-Abelian quasiparticle, the two particles will each gain a neutral
fermion \cite{SFN0502,SH0602,BKS0603,FFN0717}.  Such a non-trivial braiding is
captured in Eq. (\ref{eq:sigmafusionchannel}) because the sector of the island has been altered. Since there is no
channel in the identity representation at all three edges the expectation
value of this operator is zero.

If one calculates the tunneling conductance for the case with a
$\sigma$-quasiparticle present inside the interferometer one finds
\begin{equation} 
G_\text{tun}(T,V)=G^\text{ave}_\text{tun}(T,V),\qquad
G^\text{osc}_\text{tun}(T,V)=0.
\end{equation} 
The average tunneling conductance is still given by
Eq.~(\ref{eq:averageconductance}), but the interference term has vanished, see
Fig.~\ref{fig:mainplot}c. This follows from Eq.~(\ref{eq:sigmafusionchannel})
because the operator that caused interference before now has a zero
expectation value. An alternative explanation is that in encircling the island
a $\sigma$-quasiparticle flips its internal two-dimensional state and
`internal-spin-up' and `internal-spin-down' do not interfere. Also one can
explain the vanishing interference as being able to tell which path the
tunneling quasiparticle took, because this information can be determined from
a hypothetical measurement of the sectors of the system before and after a
tunneling event.

\section{Interference restored through $\psi$-$\psi$-tunneling} 

In the presence of a $\sigma$ quasiparticle on the central island the
interference vanished to leading order in $\Gamma_1$ and $\Gamma_2$.  Here we
will consider the situation where the the central island is close to one of
the edges (which we choose to be the left edge).  We will include a tunneling
process between the central island and the left edge. 
For simplicity we restrict ourselves to
the tunneling of the neutral fermions
only, because this process leaves the sector of the
island unaltered and furthermore this is a charge-neutral operation.
We will see that the such higher order tunneling
processes can potentially restore some interference. 

The tunneling of the neutral fermions is described by the tunneling Hamiltonian
\begin{equation} 
H_\text{tun}\to
H_\text{tun}+\Gamma_3\psi_\text{L}(t,x_3)\psi_\text{I}(t),
\end{equation} 
where the tunneling amplitude $\Gamma_3$ is real-valued to
ensure hermiticity of the Hamiltonian (operators on different edges commute).

Inclusion of this operator has the potential to restore interference because
the fusion channel of the operator 
\begin{equation}
\big[\sigma_\text{L}\sigma_\text{R}\big]^\mathbbm{1}_i\big[\psi_\text{L}\psi_\text{I}\big]^\mathbbm{1}\big[\sigma_\text{L}\sigma_\text{R}\big]_j^\mathbbm{1}\xrightarrow[\displaystyle(\sigma)_\text{I}]{\text{fus.ch.}}\mathbbm{1}_\text{L}\mathbbm{1}_\text{R}\mathbbm{1}_\text{I}+\psi_\text{L}\psi_\text{R}\mathbbm{1}_\text{I}
\end{equation} 
now has the full identity channel in it and hence a non-zero
expectation value, whereas without the additional $\psi_\text{L}\psi_\text{I}$
contribution it vanished.

We find a contribution to the tunneling conductance to leading order in
$\Gamma_3$ which contains some non-zero interference. The average value of the
conductance is still unaltered to leading order, but the oscillation amplitude now behaves
as 
\begin{equation}
G^\text{osc}_\text{tun}(T,V)=2|\Gamma_1\Gamma_2|\Gamma_3\,T^{-2}\,f_1\!\left(\frac{qV}{2\pi
T}\right).  \label{eq:Gamma3conductance}
\end{equation} 
The interference now has a different power of $T$, $T^{-2}$ instead of
$T^{-3/2}$, and a different dependence on $V/T$  through a dimensionless
function $f_1(y)$, as depicted in Fig.~\ref{fig:mainplot}d.

Since the tunneling amplitudes $\Gamma_i$ are generally considered to be
unknown, except for that they should be small, the different scaling of the
interference is the only way to tell one situation (interference with no
central $\sigma$, Fig.~\ref{fig:mainplot}b) from the other (interference with
a central $\sigma$ but also $\psi$-$\psi$-tunneling,
Fig.~\ref{fig:mainplot}d).

An observation of $\psi$-$\psi$-tunneling would indicate that the
topologically protected zero-mode space is sensitive to `spin/qubit' flips.
It also help us to understand how the destruction of interference by
non-Abelian statistics is restored as the non-Abelian particle between the the
two point contacts is moved near an edge.

\subsection{Calculation of the leading island tunneling contribution} 

The
steady state tunneling current $I_\text{tun}$ is calculated by expanding the
time evolution operator, starting from an initial state $\ket{0}$
\begin{gather} 
I_\text{tun}(t)=\bra{\varphi(t)}\, J\, \ket{\varphi(t)}\notag\\
\ket{\varphi(t)}=\mathcal{T}\{e^{-i\int^t_{-\infty}dt'[
H_0+H_\text{tun}(t')]}\}\ket{0},
\end{gather} 
where $\mathcal{T}\{\ldots\}$ indicates time-ordering. Up to second order in $H_\text{tun}$ this becomes
\begin{multline}
I_\text{tun}(t)=-i\int^t_{-\infty}\!\!\!\!\!\!dt'\,\bra{0}[J(t),H_\text{tun}(t')]\ket{0}\\
+\int^t_{-\infty}\!\!\!\!\!\!dt_1\,\int^t_{-\infty}\!\!\!\!\!\! dt_2\,\Big(
-\bra{0}\mathcal{T}\{ J(t) H_\text{tun}(t_1) H_\text{tun}(t_2)\}\ket{0}\\
\shoveright{-\bra{0}\mathcal{T}\{ J(t) H_\text{tun}(t_1)
H_\text{tun}(t_2)\}\ket{0}^*}\\ +\bra{0}H_\text{tun}(t_1)  J(t)
H_\text{tun}(t_2)\ket{0}\Big)+\ldots.  \label{eq:Itunexpansion}
\end{multline} 
The tunneling current operator $J$ is given by $iqe$ times
$H_\text{tun}$ from Eq.~(\ref{eq:Htunneling}) with appropriate minus signs in
the Hermitean conjugated part.

One term that appears in Eq.~(\ref{eq:Itunexpansion}) is the correlation
\begin{equation}
\bra{0}\mathcal{T}\{\sigma_\text{L}(t_1)\sigma_\text{L}(t_2)\sigma_\text{R}(t_1)\sigma_\text{R}(t_2)\psi_\text{L}(t_3)\psi_\text{I}(t_3)\}\ket{0}
\end{equation}
From  Eq.~(\ref{eq:sigmafusionchannel}), we see that
$\sigma_\text{L}\sigma_\text{L}\sigma_\text{R}\sigma_\text{R}$ can fuse into $\psi_\text{L}\psi_\text{I}$
if there is an $e/4$ non-Abelian quasiparticle between the two junctions.
Thus the above correlation is non-zero.

According to the conformal field theory, the time-ordered three-point-function
(at zero temperature) is given by
\begin{multline}
\bra{0}\mathcal{T}\{\sigma(t_1)\sigma(t_2)\psi(t_3)\}\ket{0}\\
=\frac{(\delta+i|t_1-t_2|)^{3/8}}{(\delta+i|t_1-t_3|)^{1/2}(\delta+i|t_2-t_3|)^{1/2}}.
\end{multline} 
Using $\bra{0}\mathcal{T}\{\sigma_\text{R}(t_1)\sigma_\text{R}(t_2)\}\ket{0}\sim
1/(\del+i|t_1-t_2|)^{1/8}
$, we find
\begin{align}
&\bra{0}\mathcal{T}\{\sigma_\text{L}(t_1)\sigma_\text{L}(t_2)\sigma_\text{R}(t_1)
\sigma_\text{R}(t_2)\psi_\text{L}(t_3)\psi_\text{I}(t_3)\}\ket{0}
\nonumber\\
&=\frac{(\delta+i|t_1-t_2|)^{1/4}}{(\delta+i|t_1-t_3|)^{1/2}
(\delta+i|t_2-t_3|)^{1/2}}
\end{align}
Here we have assumed that the correlation of $\psi_\text{I}(t_3)$ with other
operator does not depend on $t_3$.

The leading order contribution in the tunneling current is the familiar linear
response result.  But note that the next contribution contains both
time-ordered and non-time-ordered parts. The non-time-ordered part is basicly
a Keldysh contour-ordered term; to compute its expectation value we have to
analytically continue the time-ordered correlation functions. 

We would like to have an expression for any ordering of the
times $t_1$, $t_2$ and $t_3$, which we obtain by removing the absolute value
bars, 
\begin{align}
\label{eq:threepointnonordered}
&
\bra{0}\sigma_\text{L}(t_1)\sigma_\text{L}(t_2)\sigma_\text{R}(t_1)\sigma_\text{R}(t_2)\psi_\text{L}(t_3)\psi_\text{I}(t_3)\ket{0}
\nonumber\\
&=\frac{(\delta+i(t_1-t_2))^{1/4}}{(\delta+i(t_1-t_3))^{1/2}
(\delta+i(t_2-t_3))^{1/2}}
\end{align}
In these correlators $\delta$ is the short-distance cutoff and
we set the three-point-function prefactor to one.

In each term in the full expansion of Eq.~(\ref{eq:Itunexpansion}) we assume
we can decompose the expectation value of the product of operators into
correlators of the $c=1$ and $c=\frac{1}{2}$ theories for each edge
separately, which takes the typical form 
\begin{equation}
\langle(\sigma\sigma\psi)_\text{L}
(\sigma\sigma)_\text{R}\psi_\text{I}\rangle\langle
e^{\frac{i}{2}\phi}e^{-\frac{i}{2}\phi}\rangle_\text{L}\langle
e^{\frac{i}{2}\phi}e^{-\frac{i}{2}\phi}\rangle_\text{R}.
\end{equation}
The total scaling dimension of the above operators is 1. From the
scaling consideration in the $\int dt_1 \int dt_2$ integral, we find that such
an operator will contribute $\delta I_\text{tun}\propto T^{-1}$, or more
precisely
\begin{equation}
\label{dItun}
\delta I_\text{tun}=\frac{4 q e}{\pi
T}2\Re\!\left(\Gamma_1\Gamma_2^*e^{i\theta_\text{AB}}\right)\Gamma_3\,
F_1\!\left(\frac{qV}{2\pi T}\right)   .
\end{equation}

The calculation simplifies considerably because of a
cancellation: the numerator in
Eq.~(\ref{eq:threepointnonordered}) gets cancelled by the product of
the $\phi$ two-point functions. 
Correlators on the non-dispersive island edge were set equal to a constant.
Working out the double-integrals in
Eq.~(\ref{eq:Itunexpansion}) it turns out that the two time-ordered
contributions depend on the cutoff and only the non-time-ordered part
contributes in the limit of zero $\delta$.
The contribution we find to first order in $\Gamma_3$ in the tunneling
current, working at finite temperature by letting $(\delta+i t)^{g}\to (\sin[\pi
T(\delta+i t)]/\pi T)^{-g}$ and setting $x_i=0$, is Eq.~(\ref{dItun})
with $F_1$ given by
\begin{equation}
F_1(y)=\int_0^\infty\!\!\!\!\! du_1
\sin(2 y u_1)\!\!\int_0^{u_1}\!\!\!\!\!du_2\frac{1}{\sqrt{\sinh u_2
\sinh(u_1-u_2)}}.
\end{equation}

We find that we can approximate the integral over $u_2$
reasonably well with the function $\pi/\cosh\frac{u_1}{2\sqrt{2}}$.  By taking
the derivative with respect to $V$, $f_1(y)=F_1'(y)$, and absorbing constants
into the $\Gamma_i$ we arrive at the expression
Eq.~(\ref{eq:Gamma3conductance}) for the oscillation amplitude of the tunneling
conductance. Note that the $\Gamma_3$ contribution is pure interference
\emph{only}, there is no contribution to the average conductance.

\subsection{Flow to fixed point} Since the $\Gamma_3$ contribution to the
interference oscillation amplitude increases more rapidly than the average conductance as
$T$ goes to zero ($T^{-2}$ versus $T^{-3/2}$), we expect the present result to
be unstable towards lowering of the temperature. And we can ask the question
what fixed point this setup flows to; more carefully worded: by lowering the
temperature and at the same time making $\Gamma_1$ and $\Gamma_2$ smaller as
well such that tunneling between left and right edges is still weak, what
interference pattern does this flow to? Since in our setup the present $e/4$
non-Abelian quasiparticle on the island is not affected by the $\psi$-$\psi$
tunneling, the junction may flow to a non-trivial fixed point.

\section{Summary}

We calculate the scaling behavior of the non-linear I-V tunneling curves for a
two-point-contact tunneling junction between two edges of the
$\nu=\frac{5}{2}$ non-abelian fraction quantum Hall states.  Using the fusion
rule of the tunneling operators, we can calculate the non-linear I-V tunneling
curves for both cases with and without a $e/4$ non-abelian quasiparticle
between the two contacts.  It was suggested that the presence of the $e/4$
quasiparticle between the two contacts destroys the interference between the
two tunneling paths.  We show how to obtain such a result within a
quantitative dynamical edge theory.  Such a dynamical understanding allows us
to calculate how the interference reappears as the $e/4$ quasiparticle is
moved closer to an edge.  In particular, we consider the effect of a
$\psi$-$\psi$ tunneling between the island (which traps a $e/4$ quasiparticle)
and an edge.  We find that such a coupling between the island and the edge
makes the interference pattern reappear.  The scaling behavior of the
induced interference pattern is calculated as well.

\begin{acknowledgments}
We would like to thank Bertrand Halperin for fruitful discussions and for
making us aware of Ref.~\cite{HRSS07u}, in which a related problem is studied
and a slightly different result is obtained.  This research is supported by
NSF Grant No. DMR--04--33632 .
\end{acknowledgments}

\bibliography{intfnab,../../bib/wencross,../../bib/all,../../bib/publst}

\end{document}